\documentclass[useAMS,usenatbib,fleqn]{mnras}
\usepackage{calrsfs}
\DeclareMathAlphabet{\pazocal}{OMS}{zplm}{m}{n}
\usepackage{multirow}
\usepackage{hhline}
\usepackage{times}
\usepackage{graphics,epsf}
\usepackage[T1]{fontenc}
\usepackage{aecompl} 
\usepackage{amsmath}               
\usepackage{placeins}
\usepackage{booktabs}
\usepackage{amsfonts}               
\usepackage{amssymb}                
\usepackage{epsfig}                 
\usepackage{epstopdf}
\usepackage{rotating}
\usepackage{color}
\usepackage{tabularx}
\usepackage{url}

\def \apj{ApJ}

\def \aj{AJ}

\def \mnras{MNRAS}
\def \apjl{ApJ Lett.}

\def \nat{Nature}

\def \pasp{PASP}

\def \araa{ARA\&A}

\begin{document}

\title[Chaotic evolution of misaligned exomoons and Warm Jupiters]{Chaotic quadruple secular evolution and the production of misaligned exomoons and Warm Jupiters in stellar multiples }

\author[Grishin et. al.]{
Evgeni Grishin,$^{1}$
Dong Lai,$^{2}$
Hagai B. Perets $^{1}$
\\
$^{1}$Physics Department, Technion - Israel institute of Technology, Haifa,
Israel 3200002\\
$^{2}$Cornell Center for Astrophysics and Planetary Science, Department of Astronomy, Cornell University, Ithaca, NY 14853, USA\\
E-mail: eugeneg@campus.technion.ac.il (EG); dong@astro.cornell.edu (DL); hperets@physics.
technion.ac.il (HBP)}
\maketitle

\begin{abstract}
We study the chaotic and secular evolution of hierarchical quadruple systems in the $3+1$ configuration, focusing on the evolution of mutual inclination of the inner binaries as the system undergoes coupled Lidov-Kozai (LK) oscillations. We include short-range forces (SRF; such as those due to tidal and rotational distortions) that control the eccentricity excitation of the inner binary. The evolution of mutual inclination is described, a priori, by two dimensionless parameters, $\pazocal{R}_0$, the ratio between the inner and outer LK time-scales and $\epsilon_{SRF}$, the ratio between the SRF precession and the inner LK precession rates. We find that the chaotic zones for the mutual inclination depend mainly
on  $\pazocal{R}_0$, while $\epsilon_{SRF}$
controls mainly the range of eccentricity excitation. The mutual inclination evolves chaotically for $1\lesssim \pazocal{R}_0\lesssim 10$, leading to large 
misalignments. For $0.4 \lesssim \pazocal{R}_0 \lesssim 0.8$, the system could be  weakly excited and produce bimodal distribution of mutual inclination angles. Our results can be applied to  exomoons-planets in stellar binaries and Warm/Hot Jupiters in stellar triples. Such systems could develop large mutual inclination angles if the inner binary is tight enough, and also high eccentricities, depending of the strength of the short-range forces. Future detections of tilted Warm/Hot Jupiters and exomoons could put our mechanism under observational tests. 
\end{abstract}


\section{Introduction}
To date, hundreds of close (e.g. $a\lesssim 1\rm{AU}$)  giant planets have been discovered \citep{2011PASP..123..412W}, with distant companions evidence of $\gtrsim 50\%$ \citep{Kn14}. This population of close giant planets is commonly divided into Hot Jupiters (HJ) and Warm Jupiters (WJ), with orbital periods $P<10 \rm{day}$ and $P>10 \rm{day}$, respectively.  It is believed that HJs and WJs do not form in-situ, but migrate inward by various mechanisms. Disc migration invokes planet-disc interactions \citep{GT80,Tanaka02}, while high-e involved eccentricity excitation due to an additional planetary or stellar companion and tidal dissipation (e.g. \citealp{scat1, scat2, E98,E01,Wu2003,2007ApJ...669.1298F,JT08,WL11,Pet15LKHJ,P15b}; see \citealp{munoz16} for discussion and references). While HJs are mostly circular, WJs have moderate eccentricities, $e\gtrsim 0.2$ \citep{and17highe}, thus high-e migration is more prominent for WJs. 

Moons are common in the Solar System. Various surveys and detection techniques have been proposed for exomoon discoveries \citep{Kip09,S10,F17}. However, contrary to exoplanets, exomoon detections remain elusive (see \citealp{H17r} for review and references therein). Exomoon occurence is important for understanding the architecture of compact planetary systems \citep{Kane17}, early planet formation \citep{Nat15,Nat17}, and the history of circumplanetary disks \citep{CW02,ZL17}.

\begin{figure}
\begin{centering}

\includegraphics[width=7.7cm]{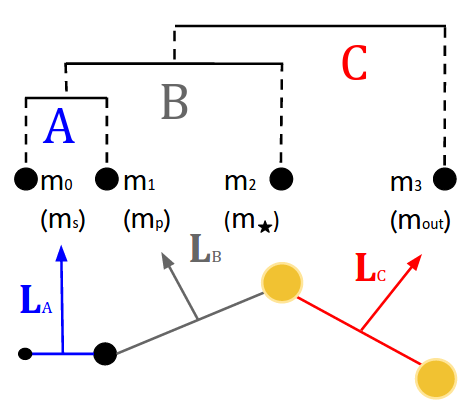}
\par\end{centering}
\caption{\label{fig1}Setup of hierarchical $3+1$ quadruple systems and their respective angular momentum vectors $\boldsymbol{L}_A, \boldsymbol{L}_B,\boldsymbol{L}_C$. A particular interest of this paper is the exo-moon problem: We consider a stellar binary (C), with a giant planet ($m_p$) orbiting the primary star ($m_\star$), and a satellite ($m_s$) orbiting the planet.}
\end{figure}

The dynamics of WJs, exomoons and other systems can be studied in the context of hierarchical systems. In triple hierarchical systems in particular, Lidov-Kozai (LK) oscillations \citep{Lidov62,Kozai62} naturally excite the inclination and eccentricity of the inner binary \citep{I97, Holman97,Ford2000,Naoz2016review}. 
The importance of triple secular dynamics has been recognised in a plethora of applications (e.g.  \citealp{1998MNRAS.300..292K,2009ApJ...697.1048P,Per+09,2012ApJ...757...27A,T16,Grishin17,LiuLai17,PA17}). 
Studies on quadruple hierarchical systems are more sparse \citep{Th13,Hamers15,HamersC,HL17,Th17}, although quadruple systems constitute a considerable fraction of hierarchical stellar systems \citep{2014AJ....147...86T,2014AJ....147...87T}.  An exomoon around a giant planet inside a
stellar binary is another likely example of quadruple system:   Figure \ref{fig1}  sketches the configuration of a $3+1$ hierarchical quadruple system. Three binaries can be identified; we refer to those with the smallest, intermediate and largest semi-major axes as binary A,B, and C respectively. 

The qualitative behaviours of quadruple systems has been described in \cite{Hamers15,HamersC,HL17}. In particular, \citep{Hamers15} numerically finds chaotic-like behaviour of the innermost binary for similar secular time-scales $t_{AB,0} \approx t_{BC,0}$ (see sec. \ref{sec:2.1} for definition and details). Similar chaotic behaviour is also found in triple systems with inner stellar spin  \citep{S14, SL15, S17}, and attributed to so called Chirikov criteria of resonance overlap \citep{Chirikov79}. In particular, synchronized circular inner binary of a quadruple system is effectively identical to the misaligned stellar spin case studied in \citep{SL15} (see sec. \ref{sec:3} for details). Thus, the semi-analytical methods of \cite{SL15} are applicable for quadruple systems with low eccentricity. Similar resonances occur when accounting for GR in triple systems leading to chaotic and resonant behavious \citep{naoz13}.

Restriction of low eccentricity naturally applies when other types of forces can affect the inner binary dynamics. Without dissipation, these forces are internal and more effective when the distances between the bodies are small (i.e. short range forces - SRFs). The main effect of SRFs is to induce pericenter and spin vector precession of the binary. This extra precession can quench the LK oscillations and controls the maximum eccentricity (e.g. \citealp{ 2007ApJ...669.1298F, LiuMunozLai15}).

In this paper we study the chaotic dynamics of hierarchical $3+1$ quadruple systems. We focus on the system with binary A being a planet and exomoon around a star (binary B), which in turn is perturbed by another stellar companion (binary C, see Fig. \ref{fig1} for sketch of the set-up). We study under what conditions an initially aligned binaries A and B will become eccentric and misaligned. Although we focus on such exomoon-planet-companion systems, the results could similarly be applied to WJ/HJ  in stellar multiple systems  \citep{HamersHJ, Hamers17b, HL17}. 

This paper is organized as follows: In sec. \ref{sec:2} we discuss the timescales and the expected dynamical behaviors of the system, as a function of the relevant dimensionless parameters. In sec. \ref{sec:3} we present the evolution of an inner circular binary. We show the similarities  of \cite{SL15} and \cite{Hamers15} by constructing bifurcation diagrams. In sec. \ref{sec:4} we extend our study to eccentric inner binaries by building 2D bifurcation diagrams. In sec. \ref{sec:5} we discuss the implications, including novel channels for production of misaligned and eccentric exo-moons in binary systems and migrating WJ/HJs in stellar triples. Finally, we summarize in sec. \ref{sec:6}.

\section{ Timescales and qualitative behavior}
\label{sec:2}
Consider a hierarchical quadruple system, as depicted in Fig. \ref{fig1}. Although our calculations can be easily adapted to general quadruple systems, we are particularly interested in the dynamics of a planet-moon system inside a stellar binary. Since planets and satellites are formed in circumstellar and circumplanetary discs, respectively, it is natural to consider an initially aligned configuration ($\boldsymbol{L}_A$ parallel to $\boldsymbol{L}_B$). However, the stellar companion can be on an inclined orbit, with a significant inclination angle between $\boldsymbol{L}_B$ and $\boldsymbol{L}_C$, $i_{BC}$. Our goal is to understand under what conditions substantial mutual inclination between $\boldsymbol{L}_A$ and $\boldsymbol{L}_B$,  $i_{AB}$, could be generated. In order to address this question, we first consider several relevant timescales/rates.  

\subsection{Timescales/Rates for Newtonian point masses}
\label{sec:2.1}
If the four bodies are pure Newtonian point masses, then the dynamical behaviour of the system is determined by the following timescales/rates: i) Precession rate $\omega_{AB}$ of binary $A$ around binary $B$ (e.g. $\boldsymbol{L}_A$ around $\boldsymbol{L}_B$) due to the external quadrupole. ii) Precession rate $\omega_{BA}$ of $\boldsymbol{L}_B$ around $\boldsymbol{L}_A$ due to the inner quadrupole of binary $A$. iii) Precession rate of $\boldsymbol{L}_B$ around $\boldsymbol{L}_C$ due to the outer quadrupole potential produced by $m_3$. The order-of-magnitude expressions for these rates are

\begin{align}
\omega_{AB,0} & =  t_{AB,0}^{-1} = n_A \frac{m_2}{m_A} \left(\frac{a_{A}}{a_{B}}\right)^{3}\label{AB_out}\\
\omega_{BA,0}& = t_{BA,0}^{-1} = n_B \frac{\mu_A}{m_A} \left(\frac{a_{A}}{a_{B}}\right)^{2} = \frac{L_A}{L_B} \omega_{AB,0}\label{AB_in}\\
\omega_{BC,0} & = t_{BC,0}^{-1} = n_B \frac{m_3}{m_B} \left(\frac{a_{B}}{a_{C}}\right)^{3} \label{BC_out}
\end{align}
where $n_A=\sqrt{Gm_A/a_{A}^{3}}$ and  $n_B=\sqrt{Gm_B/a_{B}^{3}}$ are the mean motions, $m_A = m_0+m_1$ and $m_B=m_A+m_2$ are the total masses of binaries $A$ and $B$, respectively, and $\mu_A=m_0m_1/m_A$ is the reduced mass of binary $A$. The subscript "$0$" implies that these are the rates for circular and co-planar configurations (i.e. $e_A=e_B=0$ and $i_{AB}=i_{BC}=0$).
Given these rates, the following qualitative behaviours are expected:

i) If $\omega_{BA,0}\gtrsim\omega_{BC,0}$, then LK oscillations of binary $B$ (driven by $m_3$) are quenched due to the inner quadrupole of binary $A$. For given $a_A$ and $a_C$ (and the relevant masses), the critical semi-major axis of binary $B$ (also called "Laplace radius of binary B") is  (e.g. \citealp{ML15PNAS}) 
\begin{equation}
 r_L(B) = \left( \frac{\mu_A m_B}{m_A m_3} a_A^2 a_C^3 \right)^{1/5}\label{Laplace1}
\end{equation} 
For binary $B$ with a semi-major axis $a_B \le r_L(B)$, LK oscillations will be quenched.  

In this paper we will consider the regime where $a_B\gtrsim r_L(B)$, so that binary B can undergo LK oscillations. For typical exomoon-planet-binary systems, this requires
\begin{equation}
{a}_{B}\gtrsim0.53 \left(\frac{m_0/m_1}{10^{-6}}\frac{m_2}{m_3}\right)^{1/5} 
\left(\frac{a_A}{10^6 \rm{km}}\right)^{2/5}\left(\frac{a_C}{10^3 \rm{AU}}\right)^{3/5} \rm{AU}\label{a_B estimation}
\end{equation}
where we have scaled the mass and separation of the satellite in binary $A$ using the approximate values for Ganymede, the largest and most massive moon of Jupiter.

ii) If $\omega_{BA,0}\lesssim\omega_{BC,0}$ (i.e. for systems satisfying Eq. \ref{a_B estimation}), then LK oscillations of binary $B$ occur normally with a characteristic timescale $t_{BC,0}$. In this case the fate of binary $A$ is determined by the adiabatic parameter \footnote{The parameter $\pazocal{R}_0$ is related to other adiabatic parameters for studying spin dynamics (replacing binary A by a stellar spin):  For example, \cite{SL15} use 
$ \epsilon_{AD}\equiv\left|\Omega_{pl}/\Omega_{ps}\right|_{e,\theta_{sl}=0}$
where $\Omega_{pl}$ is the precession rate of the orbital axis around the fixed outer companion, and $\Omega_{ps}$ is the stellar spin precession rate. \cite{And16} use a related parameter $\mathcal{A}$. The relations are $\mathcal{A} = (3/4)\pazocal{R}_0^{-1}$ and $\epsilon_{AD} = \pazocal{R}_0  \cos i_{BC,0}$. In detailed theoretical analysis, the different parameters have somewhat different significance (see \citealp{S17}).}  $\pazocal{R}_{0} \approx t_{AB,0}/t_{BC,0}$, i.e. ratio of secular timescales (Eq. \ref{AB_out} and \ref{BC_out}). For finite initial eccentricties, Eq. (\ref{AB_out}) should be multiplied by $(1-e_{B,0}^2)^{-3/2}$, and Eq.    (\ref{BC_out}) by $(1-e_{C,0}^2)^{-3/2}$. Thus \citep{Hamers15} 
\begin{equation}
\pazocal{R}_{0}\equiv\frac{t_{AB,0}}{t_{BC,0}}=\left(\frac{a_{B}^{3}}{a_{A}a_{C}^{2}}\right)^{3/2}
\left(\frac{m_A}{m_B}\right)^{1/2}\frac{m_{3}}{m_{2}}\left(\frac{1-e_{B,0}^{2}}{1-e_{C,0}^{2}}\right)^{3/2}\label{r_0}
\end{equation}

In this case, the following sub-regimes apply: 
 
ii-a) $\pazocal{R}_{0}\ll1$: In this "adiabatic" regime, the angular momentum vector $\boldsymbol{L}_A$ precesses rapidly around $\boldsymbol{L}_B$ and adiabatically follows
it as the latter precesses slowly around $\boldsymbol{L}_C$.  The misalignment angle $i_{AB}$ is approximately constant.

ii-b) $\pazocal{R}_{0}\gg1$: In this "non-adiabatic regime", the angular momentum vector $\boldsymbol{L}_A$ effectively precesses around $\boldsymbol{L}_C$, with $i_{AC}$ ramaining apprxoimately constant. Binary $A$ could be misaligned with binary $B$ and become eccentric. 

ii-c) $R_{0}\sim1$: In this intermediate regime, a complex chaotic evolution of $\boldsymbol{L}_A$ can occur, and binary $A$ can achieve large misalignments and eccentricities.

\begin{figure*}
\begin{centering}
\includegraphics[width=7.4cm]{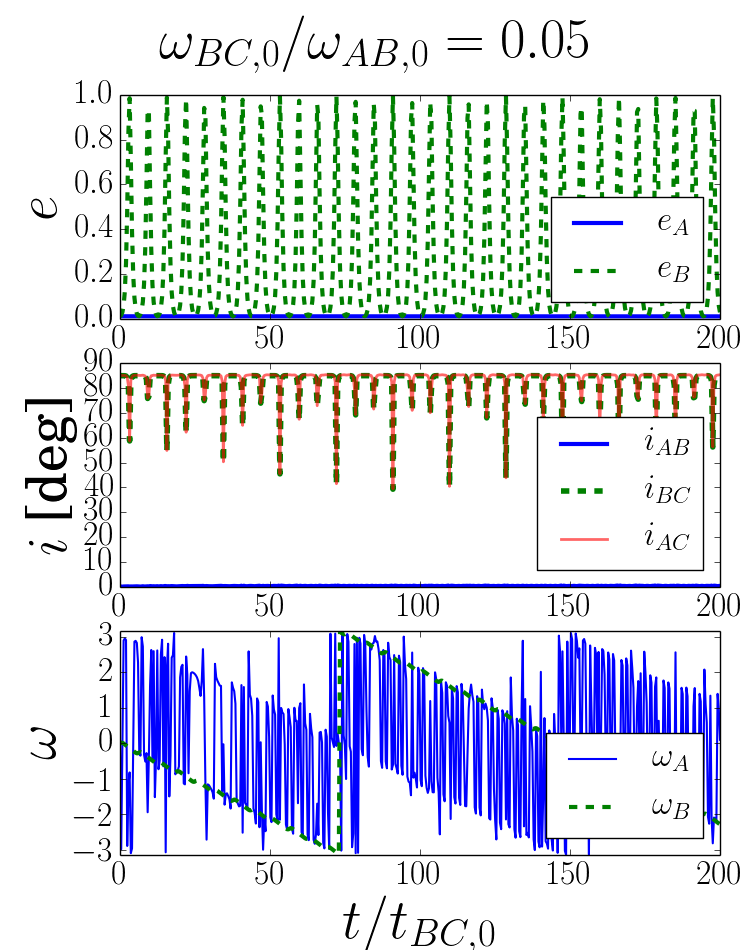}
\includegraphics[width=7.4cm]{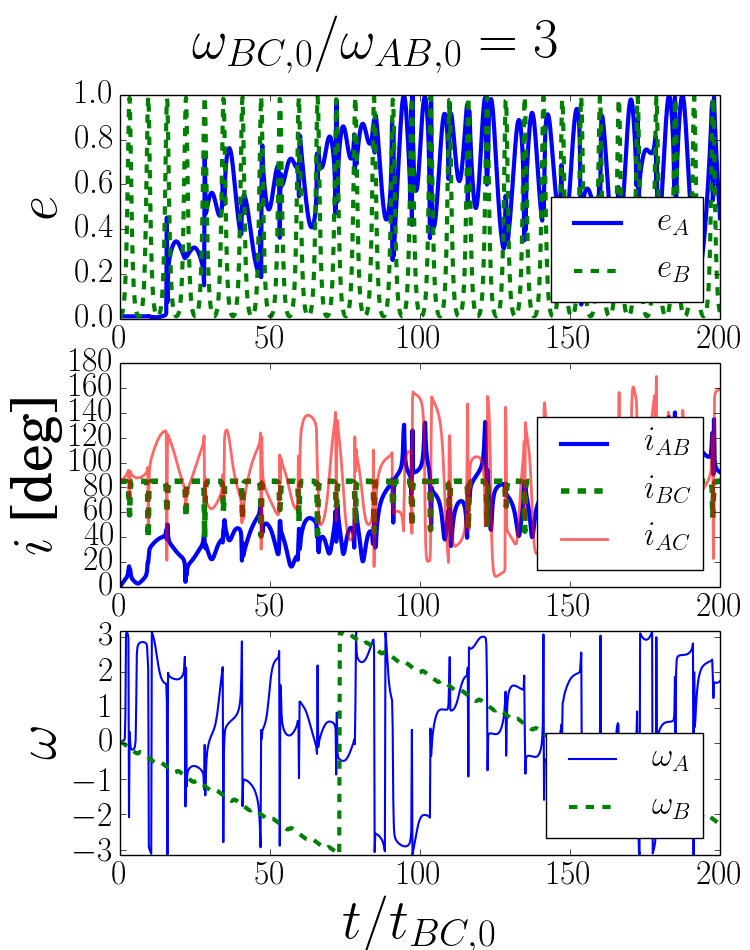}
\includegraphics[width=7.4cm]{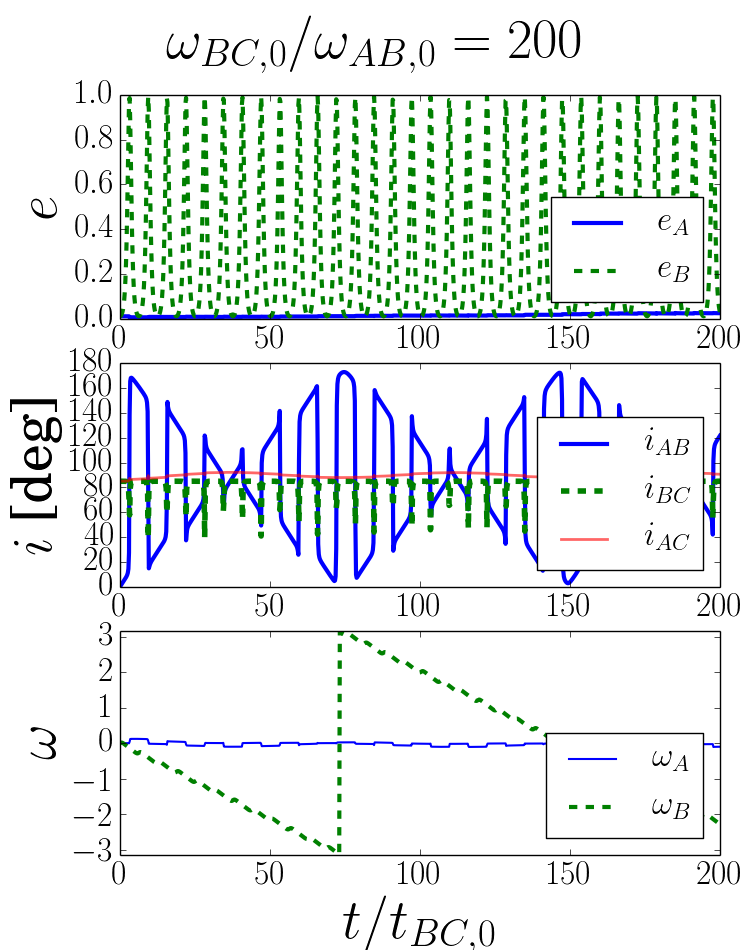}
\includegraphics[width=7.4cm]{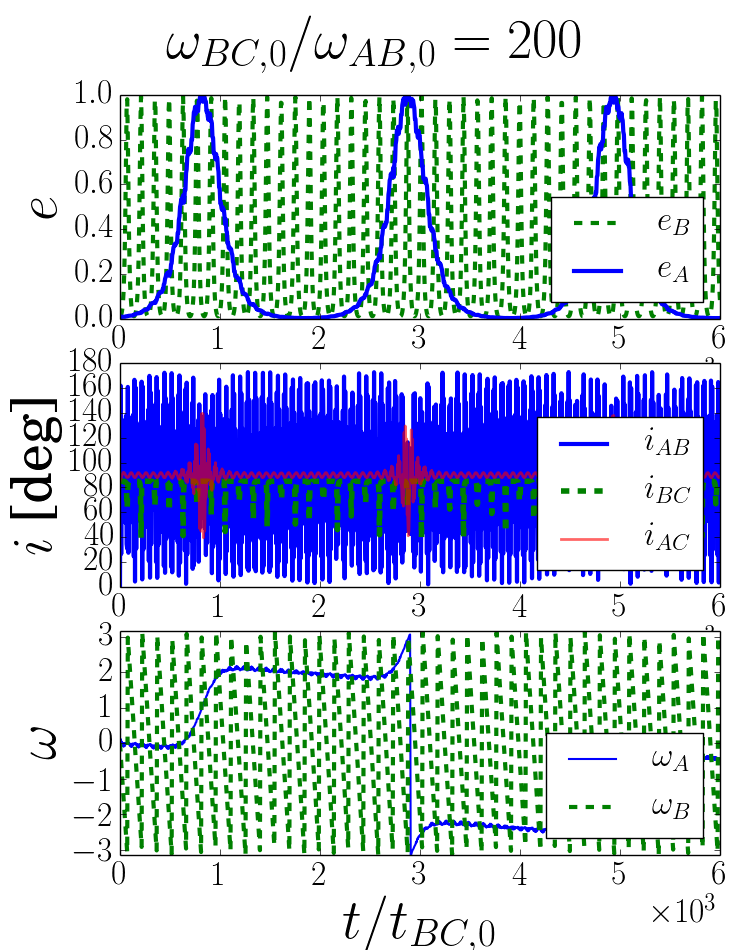}
\par\end{centering}
\caption{ \label{fig:fig2} Secular evolution of hierarchical quadruples. Each simulation starts with eccentricities $e_A=0.01$, $e_B=0.01$ and $e_C=0$, mutual inclinations $i_{AB}=0$, $i_{AC}=85^{\circ}$ and $i_{BC}=85^{\circ}$,  arguments of pericentre $\omega_A=0$, $\omega_B=0$, and $\omega_C=0$ and lines of nodes $\Omega_A=0$, $\Omega_B=0$ and $\Omega_C=0$. The systems are integrated for $200$ times of $t_{BC,0}$ (only the bottom right panel is integrated for $6000$ times $t_{BC,0}$). Four sets of figures are displayed, each showing the evolution of eccentricities, mutual inclinations and arguments of pericentre, respectively. Top left: The adiabatic regime , $\pazocal{R}_0 =0.05$.  Top left: The adiabatic regime, $\pazocal{R}_0 =0.05$. Top right: The chaotic regime, $\pazocal{R}_0 =3$. Bottom left: The non-adiabatic regime, $\pazocal{R}_0=200$. Bottom right: The same as the bottom left, only integrated to longer times. }
\end{figure*}
To demonstrate the dynamical behaviour of each regime, we evolve the secular evolution equations of the system (Appendix A) using the \textit{scipy.integrate.odeint} python package. Note that the secular approach involves averaging\footnote{The widely used term is \textit{double averaging}, but strictly speaking, the term is \textit{triple averaging} for $3+1$ quadruple systems.} over the fast angle variables on the expanded Hamiltonian, which overlooks short term corrections and dynamical instabilities\footnote{Potential consequences include: Forced eccentricities and inclinations and additional short-term corrections  \citep{Ma2001,M08, Katz16, Grishin17}, mean-motion resonances for circular orbits \citep{1999ssd..book.....M}, and quasi-resonances of true anomalies during pericenter passage for highly eccentric orbits \citep{2010ApJ...725..353D}.} \citep{Ford2000, 2012ApJ...757...27A,Hamers15}. 
  
\begin{figure*}
\begin{centering}
\includegraphics[width=8.0cm]{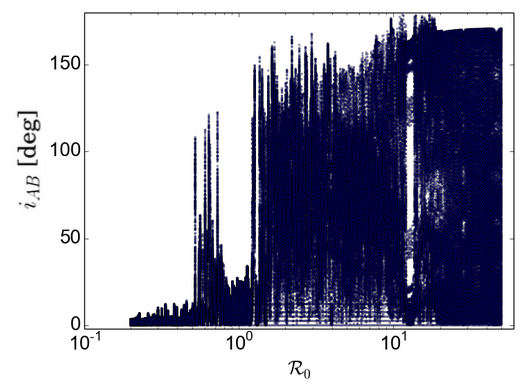}\includegraphics[width=8.0cm]{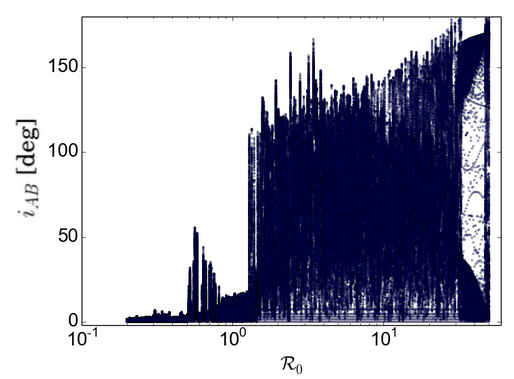}

\par\end{centering}
\caption{ \label{fig:3} Bifurcation diagrams of the misalignment angle $i_{AB}$ as a function of $\pazocal{R}_0$ (Eq. \ref{r_0}), for the circular inner binary case ($e_A=0$). Each orbit is evolved for a duration of  $500t_{BC,0}$. We record 250 different inclinations $i_{AB}$ sampled at uniform time windows and plot them in a scattered bifurcation diagram. Left panel: $i_{BC}=85^{\circ}$. Right panel: $i_{BC}=87^{\circ}$. Other initial conditions are $e_B=0.01$ and $e_A=e_C=0$; the rest are the same as in Fig. \ref{fig:fig2}.}
\end{figure*}

Figure \ref{fig:fig2} shows the secular evolution of three different systems. Each realization starts with binaries $A$ and $B$ aligned ($i_{AB}=0$), and both misaligned to binary $C$ ($i_{AC}=i_{BC}=85^{\circ}$). The top left panel shows the adiabatic regime (ii-a), where $\pazocal{R}_0=0.05$ and both $\boldsymbol{L}_A$ and $\boldsymbol{L}_B$ precess around $\boldsymbol{L}_C$, and binary $A$ remains circular and aligned with binary $B$. The top right panel is the chaotic regime (ii-c), where $\pazocal{R}_0=3$ and binary $A$ develops very high eccentricity and misalignment with either of the other binaries. The bottom left panel shows the non-adiabatic regime (ii-b), where $\pazocal{R}_0=200$ and both  $\boldsymbol{L}_A$ and $\boldsymbol{L}_B$ precess around $\boldsymbol{L}_C$. Binary $A$ exhibits LK oscillations on a time-scale $t_{AB}$. The bottom right panel shows the same run, only for $6000$ times of $t_{BC,0}$. We see that $i_{AC}$ is mostly constant, except when $e_A$ approaches unity\footnote{Generally the fluctuations in $\boldsymbol{j}_A=\sqrt{1-e_A^2}\hat{\boldsymbol{j}}_A$ over timescale $t_{BC,0}$ are small. However, when $e_A\to 1$ and $j_A \to 0$, the relative fluctuation is large which leads to large variation in $i_{AC}$, similarly to orbital flips in the eccentric LK mechanism \citep{Naoz2016review}. }. In this case the LK oscillations of binary $A$ are 'contaminated' by the precession of binary $B$ around binary $C$.    

\subsection{Short range forces} \label{sec:srf}

Short range forces (SRFs) can strongly influence the behaviour of binary $A$. SRFs are a result of internal structure of the bodies or general relativistic corrections to Newtonian gravity (e.g. \citealp{2001ApJ...562.1012E}). For conservative SRFs, the Hamiltonian approach is valid and results in precession of the pericenter $\omega$. The relevant rates are\footnote{Not to be confused with the nodal precession rates in equations \ref{AB_out}-\ref{BC_out}.} 
\begin{equation}
\dot{\omega}_{A,SRF} = \dot{\omega}_{A,rot} + \dot{\omega}_{A,tide} + \dot{\omega}_{A,GR} \label{SRF1}
\end{equation}
where the different terms account for rotational bulges, tidal bulges and GR correction, respectively. Including the internal structure of $m_1$ only, the rates are given by (e.g. \citealp{LiuMunozLai15})
\begin{equation}
\dot{\omega}_{A,SRF}=n_A \times \left\{ \begin{array}{cc}
3\frac{r_g}{a_A}\frac{1}{1-e_{A}^{2}} & \text{GR}\\
15 \frac{m_0}{m_1} \frac{k_{2,1}R_1^5}{a_A^{5}}f(e_A) & 
\text{tide}\\
\frac{3}{2} J_2 \left(\frac{R_1}{a_A}\right)^2 \frac{1}{(1-e_{A}^2)^{2}}& \text{rot}
\end{array}\right.\label{eq:srf rate}
\end{equation}
where $f(e)\equiv (1+3e^{2}/2+e^{4}/8)/(1-e^2)^5$, $r_g = Gm_A/c^2$ is the gravitational radius, $k_{2,1}$ and $R_1$ are the Love number and radius  of $m_1$, and $J_2$ is a dimensionless number related to the difference of principal inertia tensor by
\begin{equation}
(I_3-I_1)_1 = J_2 m_1 R_1^{2}\label{j2}.
\end{equation}
In the case of fast rotating planets, the most relevant SRF is the rotational bulge. When the precession time due to SRF is shorter than the secular timescale, $t_{AB,0}$, the eccentricity excitation in binary $A$ will be quenched. For initially circular binary, this occurs for $\dot{\omega}_{A,rot} \gtrsim \omega_{AB,0}$ or for 
\begin{equation}
a_A \lesssim r_L(A) = \left( J_2\frac{m_A}{m_2} R_1^{2} a_B^{3}  \right)^{1/5} \label{laplace2}
\end{equation}
where $r_L(A)$ is the planet's Laplace radius (e.g.  \citealp{T9,ZL17}).

\section{Misalignment of circular inner orbit}
\label{sec:3}

As noted in sec. \ref{sec:srf}, in the presence of a strong precession due to SRF, the orbit of binary $A$ remains circular. In this case the dynamics of $i_{AB}$ reduces to the problem of spin-orbit misalignment in three-body systems \citep{S14,SL15,And16,S17,And16_2}. The only relevant dimensionless parameter is given in Eq. (\ref{r_0}). The nodal precession rate of $\boldsymbol{L}_{A}$ around $\boldsymbol{L}_B$ is 
\begin{equation}
\frac{d\boldsymbol{\hat{L}}_A}{dt} = -\frac{3}{4} \frac{\omega_{AB,0}\cos{i_{AB}}}{(1-e_B^{2})^{3/2}} \boldsymbol{\hat{L}}_A\times{\boldsymbol{\hat{L}}}_B \label{prec1}
\end{equation}

Figure \ref{fig:3} shows the bifurcation diagram of the misalignment angle $i_{AB}$ as a function of  $\pazocal{R}_{0}$. The left and right panels show the results with outer inclinations $i_{BC}=85^{\circ}$ and $i_{BC}=87^{\circ}$, respectively. Overall, the results are consistent with Fig. 12 of \cite{SL15} and Fig. 7 of  \cite{Hamers15}. Figure \ref{fig:3} shows that while the magnitude of the chaotic excitations may depend on the initial inclination $i_{BC}$, their locations in the $\pazocal{R}_0$-space is approximately invariant. In other words, $\pazocal{R}_0$ alone determines the parameter space of chaotic excitations, but their strength depends on $i_{BC}$ as well. 

\begin{figure*}
\begin{centering}
\includegraphics[width=8.0cm]{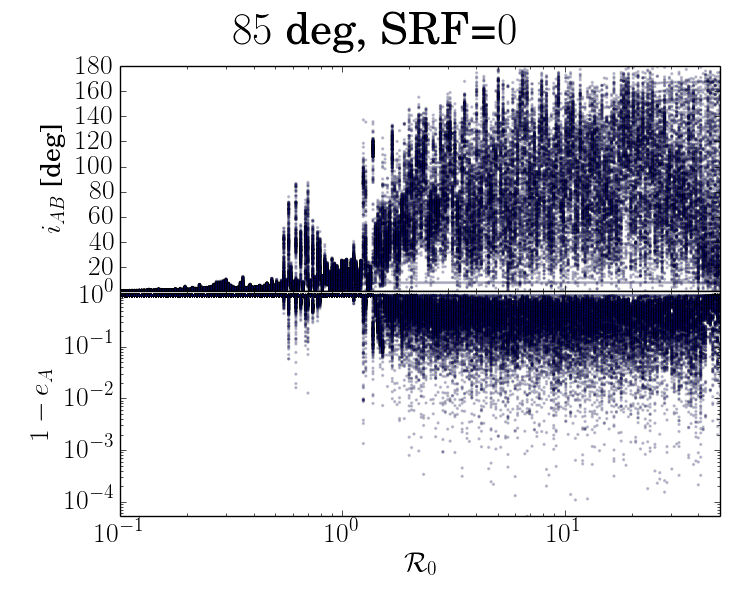}
\includegraphics[width=8.0cm]{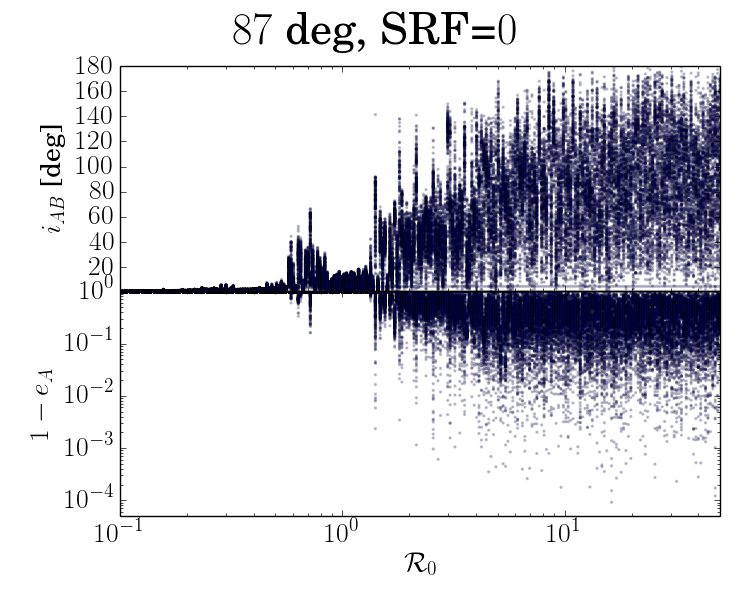}
\includegraphics[width=8.0cm]{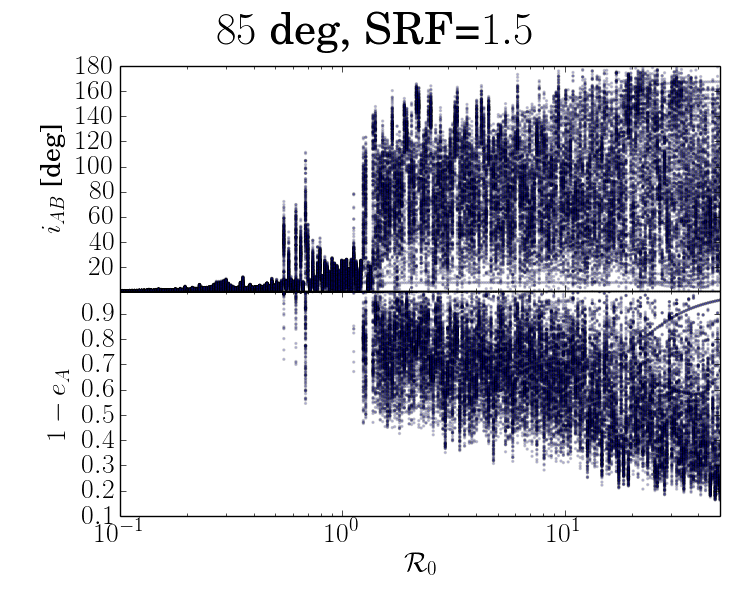}
\includegraphics[width=8.0cm]{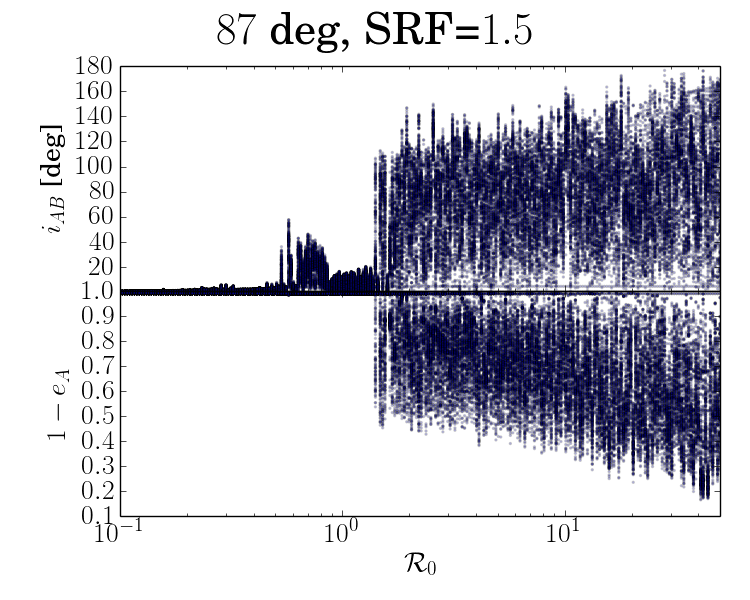}
\includegraphics[width=8.0cm]{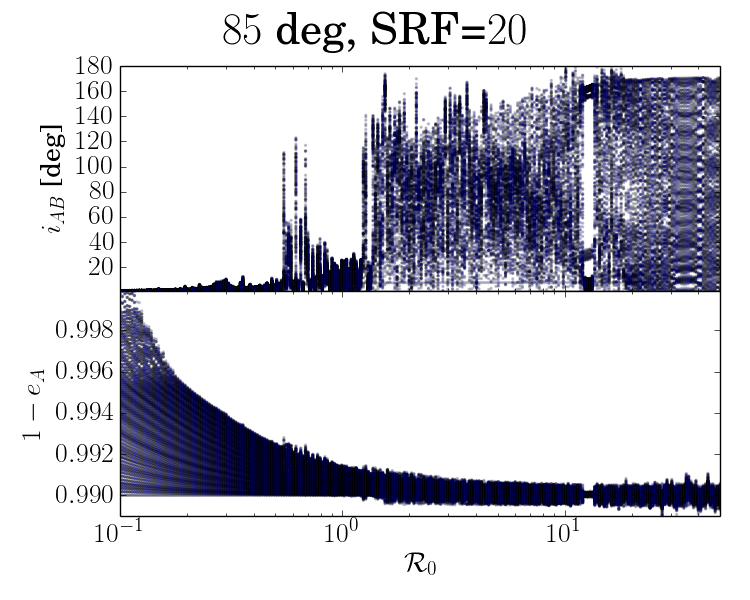}
\includegraphics[width=8.0cm]{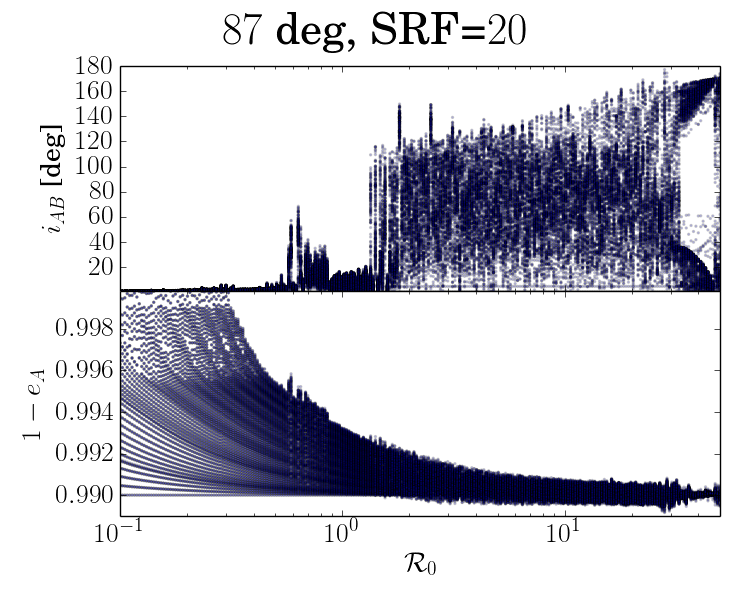}
\par\end{centering}
\caption{  \label{fig:4} Bifurcation diagrams of $i_{AB}$ as a function of $\pazocal{R}_0$, allowing for eccentricity excitation of the inner binary.  Each orbit is evolved for time $500t_{BC,0}$. We record 250 different inclinations $i_{AB}$ and plot them in a scattered bifurcation diagram. Left panels start with ${BC}=85^{\circ}$. Right panels start with $i_{BC}=87^{\circ}$. Top, middle and bottom panels are evolved with $\epsilon_{SRF}=0,1.5,20$ respectively. Other initial conditons are $e_A=0.01$, $e_B=0.01$ and $e_C=0$; the rest are the same as in Fig. \ref{fig:3}.  }
\end{figure*}

After the transition to chaos, the transition to the non-adiabatic regime ($\pazocal{R}_0 \gg 1$), where $\boldsymbol{L}_A$ effectively precesses around $\boldsymbol{L}_C$, occurs at different locations ($\pazocal{R}_0 \sim 10, 20$ for $i_{BC}=85^{\circ}, 87^{\circ}$, respectively). This transition can be seen where the maximal misalignment peak changes from sharp spikes to smooth envelope at $i_{AB,max}\approx 2i_{AB,0}$. It is because the precession rate for $\boldsymbol{L}_B$ around $\boldsymbol{L}_C$ depends also on $\cos i_{BC}$. For $i_{BC}=87^{\circ}$, this precession rate is lower and the transition to non-adiabatic regime is somewhat delayed. Note also the somewhat different quasi-periodic librating windows in the non-adiabatic regime transition.

\subsection{Effect of planet spin} \label{spin}

As discussed in sec. \ref{sec:srf} when $a_A$ lies within the Laplace radius (Eq. \ref{laplace2}), binary A will remain circular. Since the nodal precession rate of $\boldsymbol{L}_A$ around $\boldsymbol{S}_p$ is close to the apsidal precession rate $\dot{\omega}_{A,rot}$,  and the nodal precession rate of $\boldsymbol{L}_A$ around $\boldsymbol{L}_B$ is close to $\omega_{AB,0}$, the orbital angular momentum $\boldsymbol{L}_A$ is strongly tied to the planet's spin $\boldsymbol{S}_p$ for $a_A\lesssim r_L(A)$. If the spin and the orbital angular momentum are aligned, the total angular momentum is $\boldsymbol{S}_p + \boldsymbol{L}_A = (S_p+L_A)\boldsymbol{\hat{L}}_A$, binary A and the spin behave like a rigid body, and both precess around $\boldsymbol{L}_B$ with the same prescription as in Eq. (\ref{prec1}), but with the frequency $\omega_{AB,0}$ changed to 

\begin{equation}
\Omega_{s,AB} = \frac{\omega_{AB,0}L_A +2n_B^{2}(I_3-I_1)_1}{L_A + S_p}\label{spin_freq}
\end{equation}
The bifurcation diagram for this system is the same as Fig. \ref{fig:3}, with the replacement of  $\omega_{AB,0}$ by $\Omega_{s,AB}$ in the adiabatic parameter $\pazocal{R}_0$. 

Note that in the limit of $S_p \to 0$, $\Omega_{s,AB}$ reduces to $\omega_{AB,0}$, and in the limit where the spin dominates, using Eq. (\ref{j2}) we get 
\begin{equation}
\Omega_{s,AB}=\frac{2J_2}{\bar{\pazocal{C}}}\frac{n_B^{2}}{\Omega_{sp}} \label{only_spin}
\end{equation}
where $\Omega_{sp}=S_p/(I)_1$ is the spin frequency, and $\bar{\pazocal{C}}=(I)_1/m_1R_1^2$ is the moment of inertia factor (e.g. Eq. 4.113 of \citealp{1999ssd..book.....M}).

\section{Misalignment of eccentric inner orbit}
\label{sec:4}

Having understood the circular case ($e_A = 0$), we now consider the full problem, where $e_A$ is allowed to be non zero. We want to see to what extent the resulting behaviour described in sec. 3 remains valid when $e_A$ is non-zero.

For concreteness and without loss of generality, we consider only the effects of planet's rotational deformation parametrized by $J_2$. The full equations of motion are given in appendix A.

In the case of eccentric inner binary, there is additional degree of freedom, the inner binary's longitude of pericenter, $\omega_A$. The rate of precession of $\omega_A$ is driven by two processes: 

i) Precession due to outer quadrupole of binary B with rate $\omega_{AB,0}$ given in Eq. (\ref{AB_out}).

 ii) Precession rate due to the rotational bulge of the planet with rate $\dot{\omega}_{A,SRF}$ given in Eq. (\ref{eq:srf rate}).

The dimensionless parameter which determines the strength of SRF forces compared to the outer quadrupole is
\begin{equation}
\epsilon_{SRF} \equiv \frac{\dot{\omega}_{A,SRF}}{\omega_{AB,0}} \label{epsilon_srf}
\end{equation}
For $\epsilon_{SRF} \ll 1$ SRFs are negligible and $e_A$ could achieve values very close to unity, while for $\epsilon_{SRF} \gg 1$ the SRF precession completely suppresses LK oscillations and $e_A$ remains small. In the transitional region, i.e. $\epsilon_{SRF} \approx 1$ the maximal $e_A$ attained is a function of the nature of the SRF potential. Analytical expressions regarding the maximal $e_A$ in this intermediate regime can be found in \cite{LiuMunozLai15}. We use $\epsilon_{SRF}$ to control the range of $e_A$ that can be achieved.

Figure \ref{fig:4} shows the result of six different bifurcation diagrams that take different values of initial $i_{BC}$ and $\epsilon_{SRF}$. We see that indeed larger $\epsilon_{SRF}$ implies lower maximum $e_A$, and in the limit of high SRF (e.g. $\epsilon_{SRF}=20$) the structure is almost identical to the circular case in Fig. \ref{fig:3}. For lower values of $\epsilon_{SRF}$ we see that the locations of the chaotic excitations of $i_{AB}$ remains approximately the same, but the structure of the excitations is slightly different. First, it is evident that the non-adiabatic regime, $\pazocal{R}_0 \gg 1$ is devoid of quasi-periodic windows. Second, the inclinations near the chaotic transition $\pazocal{R}_0\sim1.5$ is slightly lower. It is probably because some of the transferred angular momentum goes to changing the eccentricity $e_A$ rather than the inclination $i_{AB}$. 

\begin{figure*}
\begin{centering}
\includegraphics[width=8.8cm]{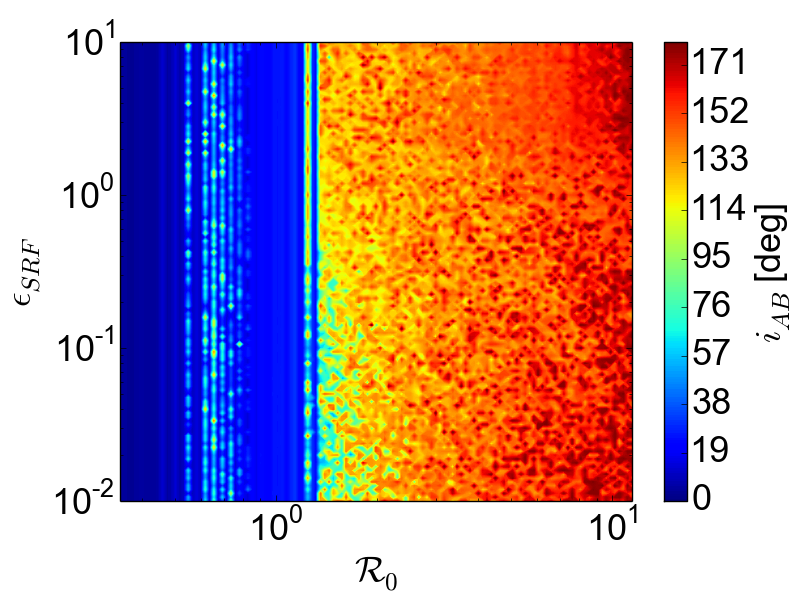}
\includegraphics[width=8.8cm]{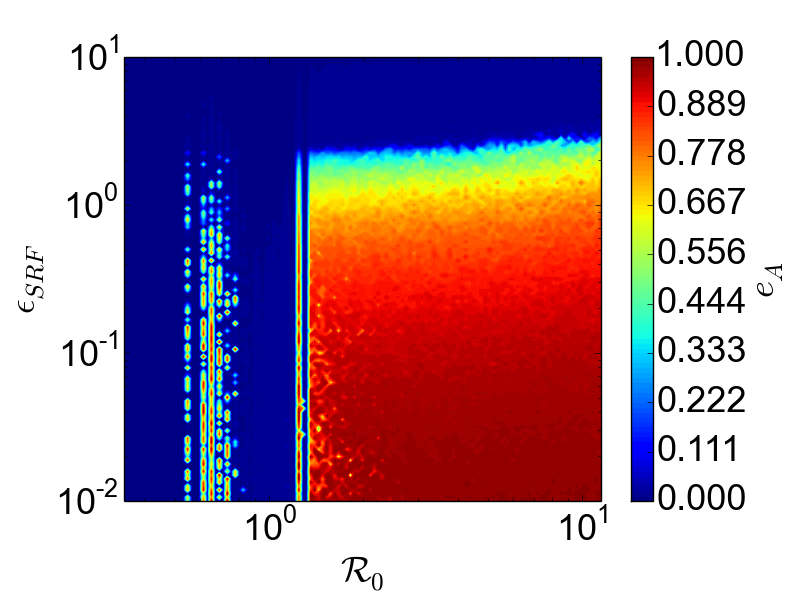}
\par\end{centering}
\caption{\label{fig:Bifurcation2d} 2D bifurcation grid in the $\pazocal{R}_0 - \epsilon_{SRF}$ space. The sampled grid points 
are log uniform from the range $\pazocal{R}_0 \in [0.3,20]$ and $\epsilon_{SRF}\in[10^{-2},10^{1}]$. The grid size is $(121,121)$ The integration time is
$t_{end}=400t_{BC,0}$. The rest of the initial conditions are the same with $i_{BC}=85^{\circ}$. Left panel shows maximal misalignment angle recorded, while right panel shows maximal eccentricity attained.}
\end{figure*}

\begin{figure*}
\begin{centering}
\includegraphics[width=8.8cm]{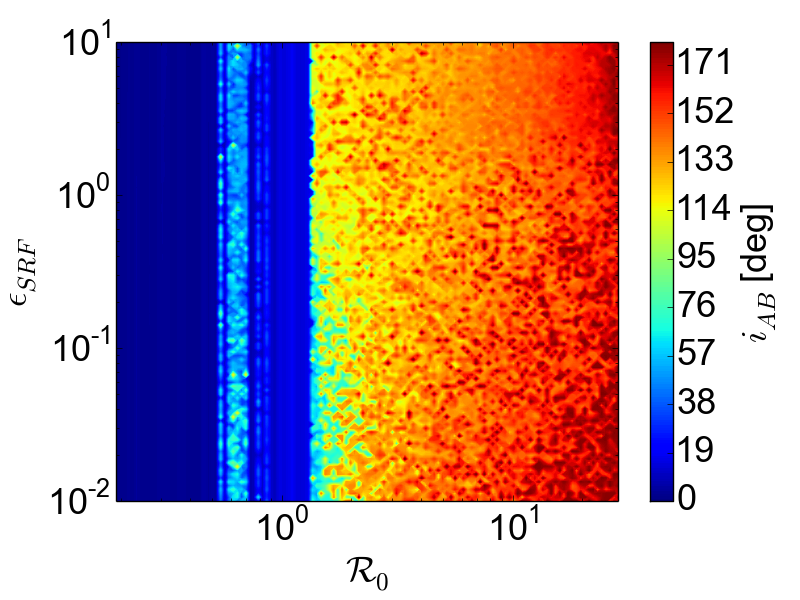}
\includegraphics[width=8.8cm, height=6.65cm]{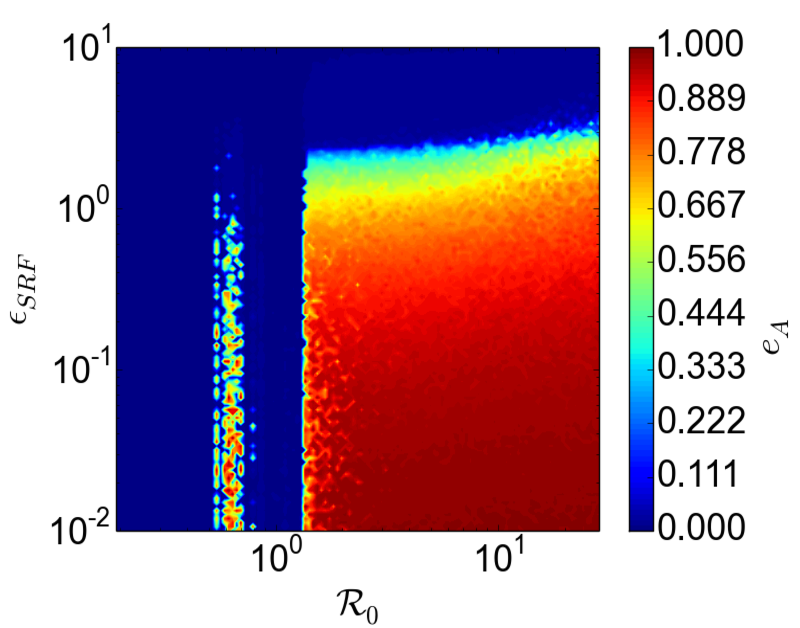}
\par\end{centering}
\caption{\label{fig:Bifurcation2d2}Same as Fig. \ref{fig:Bifurcation2d}, but with $i_{BC}=87^{\circ}$ and $\pazocal{R}_0 \in [0.2, 30]$. Note different scaling for the $x$ axis.}
\end{figure*}

In order to explore in more detail the effects of inner eccentricity on the chaotic dynamics of the misalignment $i_{AB}$, we run extensive 2D bifurcation diagrams where we vary both $\pazocal{R}_0$ and $\epsilon_{SRF}$ and record the maximal achieved inner eccentricities and inclinations. Figures \ref{fig:Bifurcation2d}  and \ref{fig:Bifurcation2d2} show the results of these 2D bifurcation diagrams for initial $i_{BC}=85^{\circ}$ and $87^{\circ}$ respectively. We see that generally the chaotic excitations of $i_{AB}$ depend mainly on $\pazocal{R}_0$ and only weakly on $\epsilon_{SRF}$. Nevertheless, we see that for small  $\epsilon _{SRF} \ll 1$, the transition to fully
chaotic misalignment is somewhat delayed (i.e., excitation of large $i_{AB}$ requires slightly larger $\pazocal{R}_0$). Finally, the top left zone in the right panels is the non-adiabatic regime, where $\boldsymbol{L}_A$ is effectively precesses around $\boldsymbol{L}_C$ (see Fig. \ref{fig:fig2}) and the maximal inclination is $i_{AB} \approx 2i_{BC,0}$. This is in contrast with the chaotic zone, where close initial conditions result in different maximal inclinations. We see that lowering the effect of SRF enhances the chaotic zone.

\subsection{Weak chaos case}
\begin{figure*}
\begin{centering}
\includegraphics[width=8.2cm]{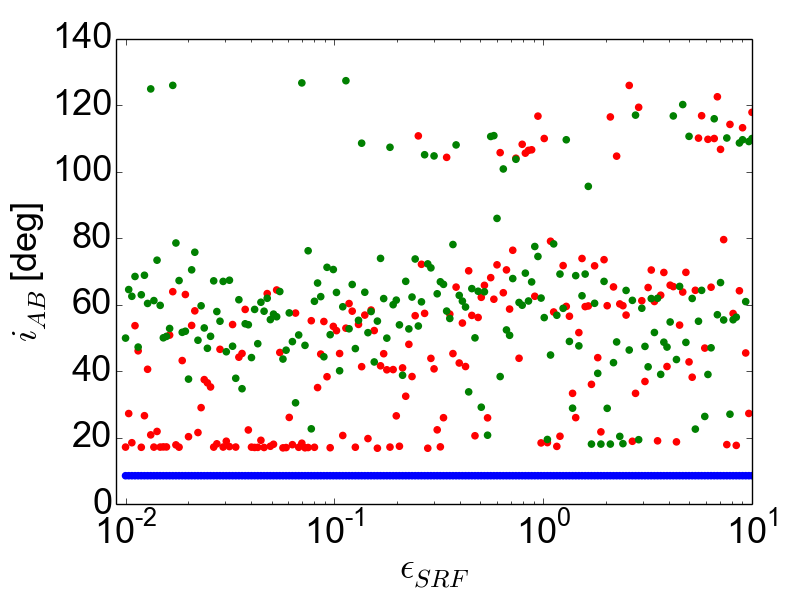}\includegraphics[width=8.2cm]{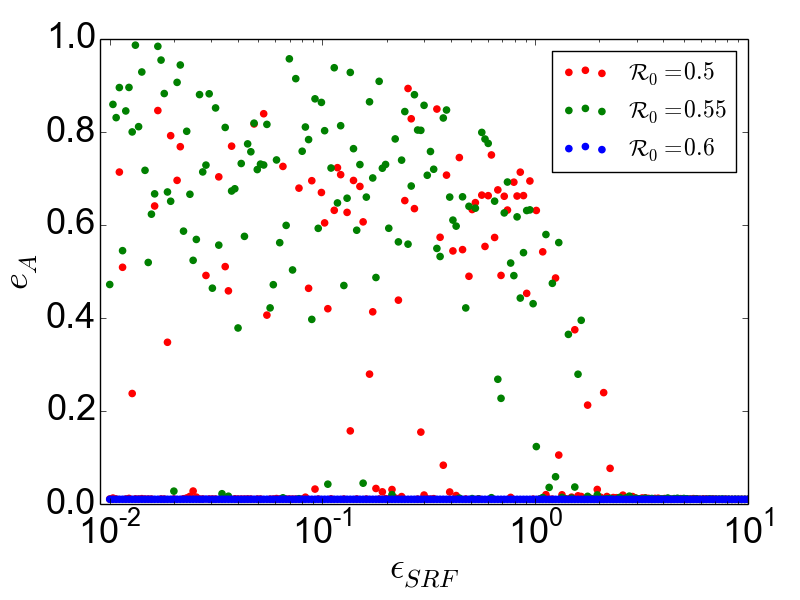}
\par\end{centering}
\caption{\label{fig:zoomin1} Distribution of  $\max(i_{AB})$ and $\max(e_{A})$
over three simulated values of $\pazocal{R}_0={0.5,0.55,0.6}$ and 200 values of $\varepsilon_{SRF}$ distributed log-uniform in $[10^{-2},10]$.} 
\end{figure*}

In addition to transition from adiabatic to chaotic regime, traces of \textit{weak chaos} are  evident for $0.4 \lesssim \pazocal{R}_0 \lesssim 0.8$, where the system exhibits chaotic excitations of $i_{AB}$ and $e_A$ (when $\epsilon_{SRF} \ll 1$), but the chaos is marginal.
\cite{SL15} explain it as weak overlapping of the $N_{max}$ resonance with
the $(N_{max}-1)$-th resonance, where $N_{max}\approx\left\lfloor 1/\epsilon_{AD}\right\rfloor $
(see their Eq. 47 and footnote 1). In our case, for given $\pazocal{R}_0$ there are many realizations with different values of $\epsilon_{SRF}$.
In order to investigate further the dependence on $\epsilon_{SRF}$
in the weak chaos region, we focus on this region and make additional zoomed-in bifurcation diagrams.

\begin{figure}
\begin{centering}
\includegraphics[width=7.4cm]{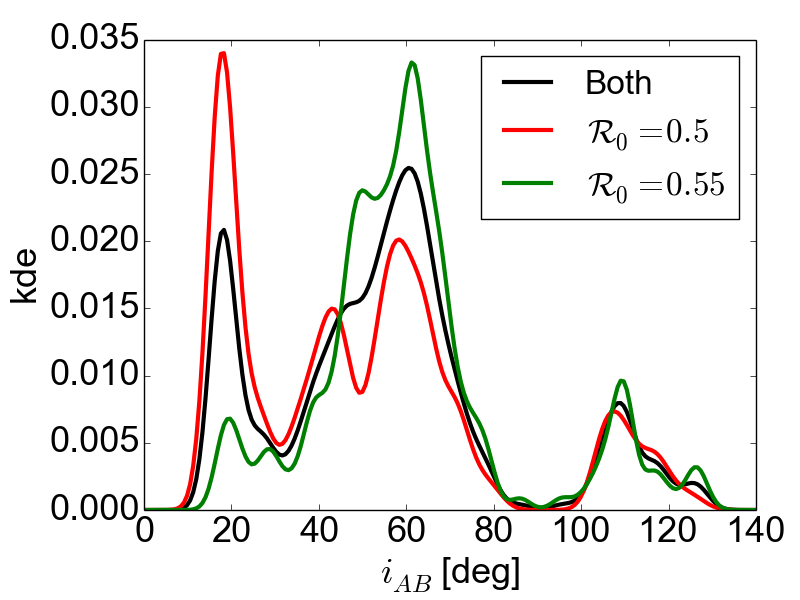}
\par\end{centering}
\caption{\label{fig:hist1} Kernel density estimation of the final inclinations for the chaotic
systems corresponding to Fig. \ref{fig:zoomin1}.}
\end{figure}

Figure \ref{fig:zoomin1} shows the results of 600 runs sampled by
$200$ values of $\epsilon_{SRF}\in[0.01,10]$ and $3$ values of
$\pazocal{R}_0=0.5, 0.55, 0.6$. For two out of three values of $\pazocal{R}_0$,
weak chaos is excited, but the final inclinations result in a bimodal distribution in $\max(i_{AB})$, regardless of $\epsilon_{SRF}$, with only few orbits achieving retrograde misalignments ($i_{AB}>90^{\circ}$).
The eccentricity, on the other hand, is distributed quasi
randomly in the allowable region. Even for $\epsilon_{SRF}\gtrsim1$
where the eccentricity is effectively quenched, the bimodal distribution of the inclination persists, and even more orbits achieve retrograde orbits. Figure \ref{fig:hist1} shows the kernel density estimation (KDE) of the chaotic orbits, confirming the bi-modality of the $\max(i_{AB})$ distribution. We use \textit{scipy.stats.gaussian\underline{\enspace}kde} python package to represent Gaussian kernels with automatic bandwidth determination.

\section{Applications}
\label{sec:5}

\begin{table*}
\begin{centering}
\begin{tabular}{|c|c|c|c|c|c|c|c|c|c|c|}
\hline 
 & $m_{0}$ & $m_{1}$ & $m_{2}$ & $m_{3}$ & $a_{A}$ & $a_{B}$ & $a_{C}$ & $R_{1}$ & $10^3 J_{2}$ & $k_{2,1}$\tabularnewline
\hline 
\hline 
S1 & $M_{\odot}$ & $M_{J}$ & $M_{J}$ & $M_{\odot}$ & $10^{-2}-10^{1}$AU & $50$AU & $2000$AU & $R_{J}$ & $14.7$ & $0.37$\tabularnewline
\hline 
S2 & $M_{\odot}$ & $M_{J}$ & $M_{\odot}$ & $M_{\odot}$ & $10^{-2}-10^{1}$AU & $200$AU & $2000$AU & $R_{J}$ & $14.7$ & $0.37$\tabularnewline
\hline 
S3 & $10^{-3}M_{J}$ & $M_{J}$ & $M_{\odot}$ & $M_{\odot}$ & $2-120$ $R_{J}$ & $10$AU & $200$AU & $R_{J}$ & $14.7$ & $0.37$ \tabularnewline
\hline 
S4 & $10^{-3}M_{\oplus}$ & $M_{\oplus}$ & $M_{\odot}$ & $M_{\odot}$ & $2-120 R_{E}$ & $10$AU & $200$AU & $R_{\oplus}$ & $1.08$ & $0.3$ \tabularnewline
\hline 
\end{tabular}
\par\end{centering}

\caption{\label{tab:ic}Hypothetical parameters of various systems. Systems S1 and S2 represent a giant planet ($m_0$) around a host star ($m_1$) with two external companions ($m_2$ and $m_3$).  Systems S3 and S4 represent an exo-moon around a giant planet  (S3) or Earth-like planet (S4) in a stellar binary. Data for taken from \citealp{And16_2, Yoder95}.}
\end{table*}

\begin{figure*}
\begin{centering}
\includegraphics[height=6.5cm]{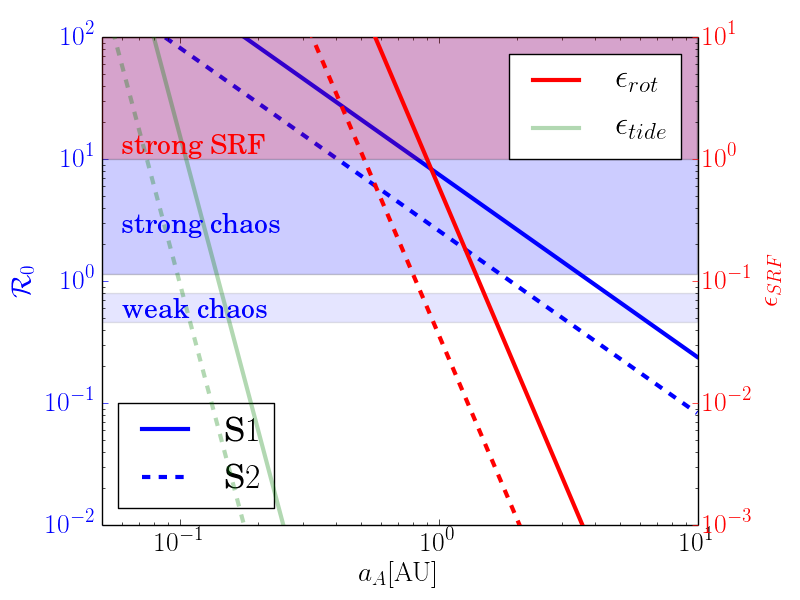}
\includegraphics[height=6.5cm]{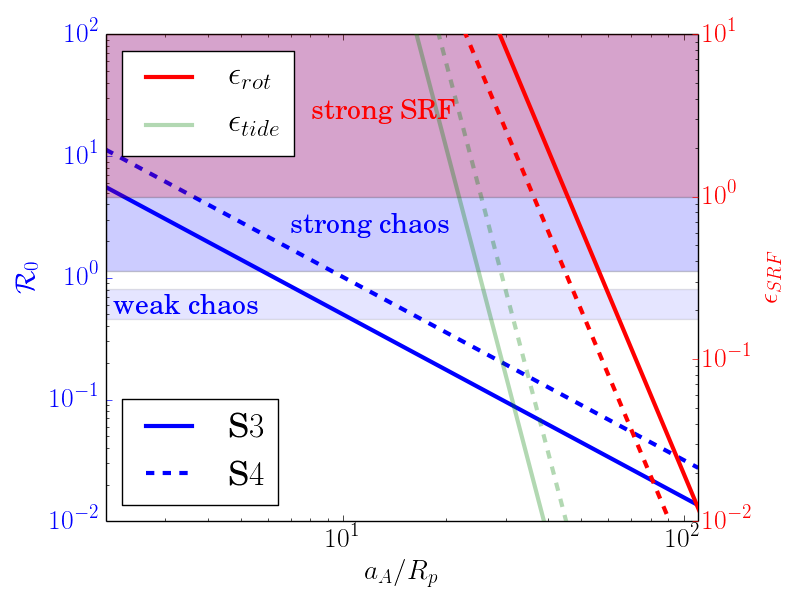}
\par\end{centering}
\caption{\label{fig:app1} Dimensionless parameters of systems S1-S4 as a function of $a_A$. Blue lines indicate the strength of the chaotic excitation, $\pazocal{R}_0$, and relate to the left y axis. Regions of weak and strong chaos are indicated by blue horizontal transparent zones. Red and green lines indicate the strength of rotation and conservative tides, respectively, and relate to the right y axis. Strong SRF are indicated by a transparent red stripe. The left panel shows systems S1 (solid) and S2 (dashed). The right panel shows S3 (solid) and S4 (dashed). In all systems the rotational bulges are stronger than tides.}
\end{figure*}

We demonstrate our results by examining possible hypothetical quadruple systems. Systems S1 and S2 describe Jupiter-like planets in a binary system with additional distant planet (S1) or in a triple stellar system (S2). Systems S3 and S4 describe exomoons of Jovian planets (S3) or Earth-like planets (S4) in binary stellar systems. For Jovian and Earth-like planets, the tidal Love number is roughly the same $k_{2,1}=0.37,0.3$, respectively \citep{And16_2, Yoder95}, and typical $J_2$ values are $1.47\cdot 10^{-2}$ and $1.08\cdot 10^{-3}$, respectively. The full list of system parameters and architecture summarized in Table \ref{tab:ic}. 

Fig. \ref{fig:app1} shows the relevant dimensionless parameters $\pazocal{R}_0$ and $\epsilon_{SRF}$ of systems S1-S4 as a function of the innermost binary separation $a_A$. In all systems, the conservative tides are negligible compared to the rotational bulges on the planet. For systems S1 and S2, SRFs are effective for $a_A\lesssim 1,0.5\rm{AU}$, respectively. It is possible that in these systems, a migrating Jupiter with $a_A\gtrsim 1\rm{AU}$ might be chaotically excited to high eccentricities and inclinations, a possible smoking gun for undetected wide companions \citep{Hamers17b}. In the weak chaos case, the bimodality of the inclination distribution could be imprinted on the final observed Hot and Warm Jupiter misalignment, as suggested by observations and detailed population synthesis studied \citep{And16, And16_2, PetHJ, HamersHJ, Hamers17b}. 

The right panel of Fig. \ref{fig:app1} shows the range of the dimensionless parameters for system S3 (solid) and S4 (dashed). The planets must be at a distance larger than a few $\rm{AU}$ (Eq. \ref{a_B estimation}) such that the distant perturber could excite the inner system via LK evolution of the planet (we took $a_B=10$AU for both systems). The induced Laplace radii are located $\sim50$ times the planetary radii, where most regular satellites lie. Inner satellites ($a_A\lesssim 10 R_p$) could undergo chaotic evolution and excite their inclination, while their eccentricity will effectively be quenched. In turn this lead to highly inclined, circular, regular satellite systems.

\section{Summary}
\label{sec:6}

We have studied the secular evolution of orbital inclinations of hierarchical quadruple $3+1$ systems. The dynamics of such systems is governed, a priori, by two dimensionless parameters ($\pazocal{R}_0$ and $\epsilon_{SRF}$; see Eqns. \ref{r_0} and \ref{epsilon_srf}). However, we found that the general behaviour of the the mutual inclination $i_{AB}$ is mostly determined by $\pazocal{R}_{0}$ (Eq. \ref{r_0}), the ratio of the secular LK oscillation time-scales of binaries AB and BC. The SRF strength, $\epsilon_{SRF}$ (Eq. \ref{epsilon_srf}) affects the eccentricity excitation of the inner binary, but plays only a minor role in the excitation of $i_{AB}$ (see Figs.5-6). We identify the following qualitative regimes.

The regular regimes include: (ii-a) the adiabatic regime, $\pazocal{R}_{0}\ll1$, where $\boldsymbol{L}_A$ precesses around $\boldsymbol{L}_B$, maintaining a constant $i_{AB}$, and (ii-b) the non-adiabatic regime, $\pazocal{R}_{0}\gg1$, where $\boldsymbol{L}_A$ precesses around $\boldsymbol{L}_C$, maintaining a constant $i_{AC}$. The chaotic regimes include: (ii-c2) the weak-chaos regime, $\pazocal{R}_{0}\in[0.4,0.8]$, where intermittent windows of order and chaos intermix, and the chaotic excitations are somewhat limited, and (ii-c1) the chaotic regime, $1\lesssim \pazocal{R}_0\lesssim 10$, where the inclination evolution is complex.

Our results are generally compatible  with \citealp{SL15, Hamers15}. Thus, the chaotic region in \citealp{Hamers15} (e.g. their Fig. 7) is indeed caused by resonance overlap. For strictly circular orbits, the case essentially the same as in \citealp{SL15}, where the stellar spin is replaced by an inner circular binary.

Allowing the inner binary eccentricity to grow does not affect the locations of chaotic regions in terms of the $\pazocal{R}_0$, but it does affect the final maximal inclinations. A weak SRF somewhat decreases the mean maximal inclination (Fig. \ref{fig:Bifurcation2d} and \ref{fig:Bifurcation2d2}) and the onset of non-adiabatic regime. In the weak chaos regime different SRFs drastically alter the inclination distribution to bimodal, regardless of the strength of SRFs. This bimodal distribution may have an imprint on the final misalignment of migrating Jupiters and exomoons in binary systems, regardless of their eccentricities.

Possible implications include the production of misaligned Warm/Hot Jupiters (see Fig. 9, left panel) with the aid of an unseen wide second planet (system S1), or a wide tertiary stellar companion (system S2). In these cases, chaotic evolution could pump the eccentricity and inclination of the migrating Jupiter, much before SRFs are effective.

More importanty, if the inner binary is a planet and an exomoon,  misaligned regular exo-moons in binary systems (S3 and S4; (see Fig. 9, right panel) could be produced by our mechanism. In these cases, the exo-moons are well within the Laplace radius of the planet, hence they remain circular, but their misalignment angle evolves chaotically. Potential discoveries of tilted exomoons could put our mechanism into observational tests.  
\\[0\baselineskip]

EG thanks the Cornell Astronomy and Space Sciences Department for their kind hospitality during July-August 2016. EG and HBP acknowledge
support from the Israel-US bi-national science foundation, BSF grant
number 2012384, European union career integration grant \textquotedblleft GRAND,\textquotedblright{}
the Minerva center for life under extreme planetary conditions and
the Israel science foundation excellence center I-CORE grant 1829/12. DL acknowledges support from NASA grant NNX14AP31G
and NSF grant AST-1715246.

\subsection*{APPENDIX A. Equations of Motion}
\label{sec:appendix}
We use the notation of \cite{Hamers15} and the evolve the vector pairs $\{\boldsymbol{j}_{A,B,C},\boldsymbol{e}_{A,B,C}\}$, where $\boldsymbol{e}_i=e_i\hat{\boldsymbol{e}}_i$ and $\boldsymbol{j}_i=\sqrt{1-e_i^{2}}\hat{\boldsymbol{j}_i}$ are the eccentricity and normalized angular momenta vectors, respectively, for binary $i={\{A,B,C}\}$. The vectors have the orthogonal relations $\boldsymbol{e}_{i}\cdot\boldsymbol{j}_{i}=0$
and $e_{i}^{2}+j_{i}^{2}=1$. Thus, there are 4 d.o.f which could be recognised by the  orbital elements
$(e_{A},\Omega_{A},\omega_{A},i_{A})$. Indeed, in a Cartesian coordinate system, the vectors could be expressed in terms of the orbital elements as
\begin{equation}
\hat{\boldsymbol{e}}_{A}  =  \begin{pmatrix}\cos\Omega_{A}\cos\omega_{A}-\sin\Omega_{A}\sin\omega_{A}\cos i_{A}\\
\sin\Omega_{A}\cos\omega_{A}+\cos\Omega_{A}\sin\omega_{A}\cos i_{A}\\
\sin\omega_{A}\sin i_{A}
\end{pmatrix} \label{e_vec}
\end{equation}

\begin{equation}
\hat{\boldsymbol{j}}_{A}  =  \begin{pmatrix}\sin\Omega_{A}\sin i_{A}\\
-\cos\Omega_{A}\sin i_{A}\\
\cos i_{A}
\end{pmatrix} \label{j_vec}
\end{equation}

The secular evolution is given by the triple-averaged quadrupole Hamiltonian (see appendix of \citealp{Hamers15}) and by the SRFs (e.g. \citealp{LiuMunozLai15})
\begin{equation}
\frac{d\boldsymbol{e}_A}{dt}  =  \left. \frac{d\boldsymbol{e}_A}{dt}\right|_{sec,AB} + \left. \frac{d\boldsymbol{e}_A}{dt}\right|_{SRF}
\end{equation} 
\begin{equation}
\frac{d\boldsymbol{j}_A}{dt}  =  \left. \frac{d\boldsymbol{j}_A}{dt}\right|_{sec,AB}
\label{eom1}
\end{equation}
The secular evolution is given by (e.g. \citealp{T9})
\begin{equation}
\label{eq:djadt}
\left. \frac{d\boldsymbol{j}_A}{dt}\right|_{sec,AB}=\frac{3\omega_{
AB,0}}{4j_{B}^{3}}\left\{ (\hat{\boldsymbol{j}}_{B}\cdot\boldsymbol{j}_{A})\boldsymbol{j}_{A}-5(\hat{\boldsymbol{j}}_{B}\cdot\boldsymbol{e}_{A})\boldsymbol{e}_{A}\right\} \times\hat{\boldsymbol{j}}_{B}
\end{equation}
\begin{equation}
\begin{aligned}
\label{eq:deadt}  
\left. \frac{d\boldsymbol{e}_A}{dt}\right|_{sec,AB} & =\frac{3\omega_{AB,0}}{4j_{B}^{3}}\left\{ (\hat{\boldsymbol{j}}_{B}\cdot\boldsymbol{j}_{A})(\boldsymbol{e}_{A}\times\boldsymbol{j}_{A})\right.\\
 & -5(\hat{\boldsymbol{j}}_{B}\cdot\boldsymbol{e}_{A})(\boldsymbol{j}_{A}\times\hat{\boldsymbol{j}}_{B}) +\left.2(\boldsymbol{j}_{A}\times\boldsymbol{e}_{A})\right\}
\end{aligned}
\end{equation}
Equations (\ref{eq:djadt}) and (\ref{eq:deadt}) are the vector form of the standard LK equations. 

The secular evolution of $\boldsymbol{e}_B$, $\boldsymbol{j}_B$ is similar, only with the replacement $A\to B$ and $B\to C$. 

The evolution due to SRF results only in the precession of the eccentricity vector, namely 
\begin{equation}
\begin{aligned}
\left.\frac{d\boldsymbol{e}_{A}}{dt}\right|_{SRF}=-\dot{\omega}_{SRF}\boldsymbol{e}_{A}\times\hat{\boldsymbol{j}}_{A}
\end{aligned}
\end{equation}
where $\dot{\omega}_{SRF}$ is defined in Eq. (\ref{SRF1}).

\bibliographystyle{mnras}

\end{document}